# Chem2Bio2RDF:
# A Linked Open Data Portal for Chemical Biology

Bin Chen[1]   Ying Ding[2]   Huijun Wang[1]
David J Wild[1]   Xiao Dong[1]   Yuyin Sun[2]
Qian Zhu[1]   Madhuvanthi Sankaranarayanan[1]

[1]School of Informatics and Computing, Indiana University, Bloomington, IN, USA
[2]School of Library and Information Science, Indiana University, Bloomington, IN, USA
{binchen|dingying| huiwang|djwild|xdong|yuysun|qianzhu|madhsank}@indiana.edu

## ABSTRACT

The Chem2Bio2RDF portal is a Linked Open Data (LOD) portal for systems chemical biology aiming for facilitating drug discovery. It converts around 25 different datasets on genes, compounds, drugs, pathways, side effects, diseases, and MEDLINE/PubMed documents into RDF triples and links them to other LOD bubbles, such as Bio2RDF, LODD and DBPedia. The portal is based on D2R server and provides a SPARQL endpoint, but adds on few unique features like RDF faceted browser, user-friendly SPARQL query generator, MEDLINE/PubMed cross validation service, and Cytoscape visualization plugin. Three use cases demonstrate the functionality and usability of this portal.

## Categories and Subject Descriptors

D.2.12 [Interoperability]: Data Mapping

## General Terms

Experimentation

## 1. INTRODUCTION

Curing disease and providing better health care require a synthesis of data and understanding across different disciplines, domains, and applications. The information and data are most useful when they are well integrated with each other. Traditional data integration techniques relying on static mapping methodologies are unlikely to be scalable. Recent advances in biology and medicine have led to an explosion of new data sources, such as genes, proteins, genetic variations, chemical compounds, diseases and drugs. Since these data are disconnected, it is very challenging for biomedical scientists to identify related genes through different diseases, to discovery new drugs, and to interconnect compounds with pathways. The recent development of chemical and biological sciences has resulted in the emergence of fields like systems biology which adopt a comprehensive approach to the study of biological systems, chemogenomics, and systems chemical biology [20].

Systems chemical biology is a new and developing area, with current application in polypharmacology [5, 12] and adverse drug reaction [22] addressing problems in efficacy and toxicity, which are the two main reasons accounting for the drug failure in the drug discovery. Polypharmacology and adverse drug reaction involve linking heterogeneous chemical and biological data with broad range of scales from small molecules (small compound, drug), to super-molecules (gene, protein), to biological systems (protein complex, pathway), and to phenotypes (disease, side effects). In addition, many databases cover similar data (called homogeneous data here) but with slightly different focuses. For instance, DrugBank[1] has drug target association, while PharmGKB[2] has similar information from different perspectives. All the heterogeneous and homogeneous data are scattered around the web and published in diverse formats (i.e., text file, XML and relational database). Data integration is essential to systems chemical biology.

Semantic Web technologies enable data integration and data interlinking on the Web and demonstrate promising potentials in life sciences, healthcare and drug discovery [17, 19, 23]. Integrating heterogeneous data is obviously necessary for advanced network biology and network medicine research [6]. Bio2RDF [3] manages to integrate public bioinformatics databases and convert them into 46 million RDF triples. Linking Open Drug Data (LODD) [13] links various sources of drug data together to answer interesting scientific and business questions. Bio2RDF already covers most of biological data (e.g., protein and pathway) and LODD has collected various sources relating to chemicals (particularly drugs), although some efforts have been made to link both together [7], its practical application is still not fully explored. The integration of chemical and biology data is actually an interaction between small compounds and proteins, usually called chemogenomics data. However, little effort has been allocated currently to integrate them. In this paper, we build the system called Chem2Bio2RDF to address these issues around system chemical biology [28].

The contribution of this paper can be summarized as follows:

- We have aggregated and converted a variety of public chemogenomics data distributed around the Web into RDF formats, which enables linking with other biological Semantic Web information resources such as Bio2RDF and LODD.
- We built up a portal called Chem2Bio2RDF and designed a faceted browser to display DrugBank data, which can be further extended to other LOD bubbles.
- We prototyped an automatic SPARQL query generator so that user can avoid writing complicated SPARQL queries.
- We provided a MEDLINE/PubMed literature cross-validation service so that the SPARQL query results can be cross-validated by related literatures in MEDLINE/PubMed.
- We prototype a Cytoscape plugin for this portal.

This paper is organized as the following: Section 2 surveys the related works; Section 3 describes the Chem2BIO2RDF portal and its unique features; Section 4 discusses three cases to prove the concept and Section 5 evaluates the methods; Section 6 makes the conclusion and points out future research.

---

[1] http://drugbank.ca/
[2] http://www.pharmgkb.org/

## 2. Related works

In 2009, there are around 1,170 molecular biology databases reported by Nucleic Acids Research (NAR). This huge amount of databases creates data deluge that bioinformaticians have huge difficulty to deal with. Most of the data only begins to make sense when it is integrated with others. W3C formed the Health Care and Life Sciences (HCLS) Interest Group to develop, advocate for, and support the user of Semantic Web technologies for biological science, translational medicine and health care. Its recent effort on linking open drug data (LODD) LODD linked RDF data from the Linked Clinical Trials dataset derived from ClinicialTriasl.gov, DrugBank (a repository of almost 5000 FDA-approved drugs), DailyMed published by the National Library of Medicine, SIDER containing information on marketed drugs and their recorded adverse reactions, TCMGeneDIT about Traditional Chinese Medicine, Diseasome with 4,300 disorders and genes from Online Mendelian Inheritance in Man (OMIM). Overall, there are more than 8.4 million RDF triples and 388,000 links to external data sources [13].

Bio2RDF [3] is a system based on RDFizer, the Sesame open source triple store and an OWL ontology. It converts KEGG, PDB, MGI, HGNC and several NCBI databases into RDF triples. It successfully applied the Semantic Web technology to convert public biomedical data into a knowledge space of RDF documents based on a shared common ontology. It demos the potential to explore the implication of four transcription factor genes in Pakinson's disease. Similar earlier efforts are YeastHub [8], LinkHub [25], BioDash [18] and BioGateway [2].

Villanueva-Rosales, Osbahr and Dumontier (2007) [25] extended the Open Biomedical Ontologies (OBO) to represent knowledge in the Saccharomyces Genomes Database so that semantically sophisticated queries can be answered by a reasoner. The authors also identified significant challenges towards such directions, such as the automated creation and population of ontologies, the efficient storage of ontological data for reasoning and the development of intuitive interfaces for others.

Cheung, et al [7] created two health care and life science semantic knowledge bases and used Semantic Web technologies to represent, map and dynamically query multiple datasets. They developed a prototype receptor explorer using OWL mappings to provide an integrated list of receptors and executing individual queries against different SPARQL endpoints. Their toolset called "FeDeRate" enables a global SPARQL query to be decomposed into subqueries against the remote databases offering either SPARQL or SQL query interfaces.

Flexible and efficient data integration is highly desired in life sciences. In summary, there are two major kinds of efforts: integrating different databases by using bridge tables or splitting queries into sub-queries to federate results. The first effort is not scalable once more tables need to be integrated as data flow programs need to be recoded. The second effort faces the major issue of query efficiency and query repackaging [7]. Semantic web approaches convert the current life science databases into a global database with unique and dereferencable URIs, machine-processable semantics, and reasoning and pattern-matching based SPARQL queries. A dereferenceable URI helps to not only uniquely identify information but also locate information. This allows anyone to create new RDF graphs by interlinking different existing RDF graphs.

## 3. Chem2Bio2RDF portal
### 3.1 Datasets

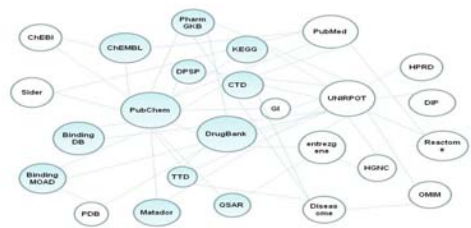

Figure 1. Overview of the Chem2Bio2RDF Datasets

The current Chem2Bio2RDF portal covers 25 datasets relating to systems chemical biology, which is grouped into 6 domains, namely chemical (PubChem Compound, ChEBI, PDB Ligand), chemogenomics (KEGG Ligand, CTD Chemical, BindingDB, MATADOR, PubChem BioAssay, QSAR, TTD, DrugBank, ChEMBL, Binding MOAD, PDSP, PharmGKB), biological (UNIPROT, HGNC, PDB, GI), systems (KEGG Pathway, Reactome, PPI, DIP), phenotype (OMIM, Diseasome, SIDER, CTD diseases) and literature (MEDLINE/PubMed). Some of data sources cross multiple domains. As the time of writing, the numbers of triples is about 78 million (Figure 1).

As the RDF presentations of the dataset in other domains have been reported (i.e., LODD, Bio2RDF), our focus is on the chemogenomics data highlighted in figure 1. Some of the major datasets we integrated are:

- PubChem BioAssay [29]: PubChem is a public repository of chemical information including structures of small molecules and various molecular properties. It has three databases namely *Compound*, *Substance* and *BioAssay*. The BioAssay database contains experimental results of the compounds in PubChem that have been tested in MLI screening centers or elsewhere against particular biological targets. We only selected the enzymatic assays that study one target and its associated active compounds.
- KEGG [30]: A collection of online databases curate chemical, genome and pathway information. One of its databases KEGG Ligand has the biological molecules associated with its enzymes.
- CTD [31]: Comparative Toxicogenomics Database (CTD) focuses on the effects of environmental chemicals on human disease.
- BindingDB [32]: It provides binding affinities on the interactions of protein with small, drug-like molecules. The data are extracted from the scientific literature, focusing on the proteins that are drug-targets or candidate drug-targets.
- Matador [33]: Compared to DrugBank, MATADOR (Manually Annotated Targets and Drugs Online Resource) provides the indirect interaction between chemicals and targets that is collected by automated text-mining followed by manual curation.
- QSAR: We manually collected a set of small QSAR datasets from two websites (http://www.cheminformatics.org/, http://w*ww.qsarworld.com/qs*ar-da*tasets.p*hp), which aim to accumulate all the QSAR dataset published along with the paper. Only the dataset with a particular target was selected.
- TTD [34]: Therapeutic Target Database (TTD) provides information about the known therapeutic proteins and nucleic acids described in the literature and their corresponding drugs/ligands.

- DrugBank [35]: It combines detailed drug (i.e. chemical, pharmacological and pharmaceutical) data with comprehensive drug target (i.e. sequence, structure, and pathway) information. The database contains nearly 4,800 drug entries and more than 2,500 non-redundant protein.
- ChEMBL [36]: ChEMBL focuses on the interactions and functional effects of small molecules binding to their macromolecular targets. It provides 500,000 bioactive compounds, their quantitative properties and bioactivities (binding constants, pharmacology and ADMET, etc).
- PDSP [37]: PDSP Ki database provides information on the abilities of drugs to interact with an expanding number of molecular targets. It provides Ki value (one measure of binding affinity) for its 766,000 interaction.
- PharmGKB [38]: It curates primary genotype and phenotype data, annotate gene variants and gene-drug-disease relationships via literature review.
- Binding MOAD [39]: Mother Of ALL Databases (MOAD) collects all well resolved protein crystal structures with clearly identified biologically relevant ligands annotated with experimentally determined binding data extracted from literature. All the structures are extracted from Protein Data Bank (PDB), containing very high-quality ligand-protein binding.

The detailed data processing was reported in our previous work [28]. We add simple provenance (What, When, Where, Why, and Who) to the RDF resources, which can be viewed at our website (chem2bio2rdf.org/datasets.html). The user can explore the RDF resources using the filter function. For example, the Figure 2 shows all the chemogenomics RDF resources which are extracted from literature.

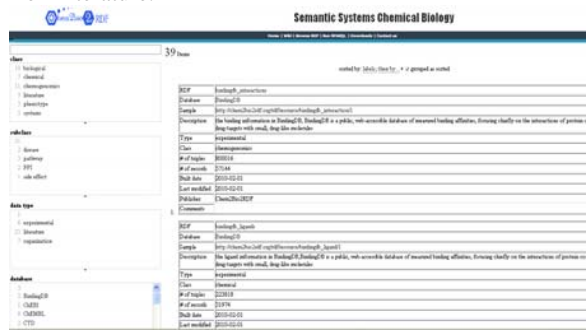

Figure 2. Faceted Browser for the datasets

All the data was processed from its original dataset, however, some of them does not meet the further utilization. For example, BindingDB uses string (i.e., >0.5) instead of number to present its binding affinity, if the user wants to select the data greater than 0.4, this text does not support this kind of search, so we have to manually separated the text into two parts, namely operator and number. In addition to data quality, the missed links between different RDF resources is a serious issue. For instance, many data formats (SMILES, SDF, InChi and CID number) are available to present one chemical, but the machine does not know they actually talk the same subject. Thus we assigned the chemical a CID number (PubChem Compound ID) if available, as PubChem is a hub of the public compounds. Meanwhile, proteins can be presented as GI number, UNIPROT ID, EC number, PDB ID and gene symbol, so we added many separated RDF resource to connect them together. GI2UNIPROT offers the conversion between GI number and UNIPROT ID. All the proteins could be eventually converted to UNIPROT ID, that allows to link to pathways, protein-protein interaction and diseases.

Chem2Bio2RDF aims to bridge chemical and biological data, we linked our data to LODD and Bio2RDF using owl:sameAs. As LODD and BioRDF have strict namespace definition and dereferenceable URI. For instance, the URI of a drug in Bio2RDF is http://bio2rdf.org/drugbank_drugs: followed by drug id. This allows us simply link our data (i.e, drug, enzyme, protein, gene, pubmed) to them.

## 3.2 Chem2Bio2RDF Faceted Browser

MIT Simile project toolsets are used to provide user-friendly faceted RDF triple browser. Figure 3 shows the difference between normal SPARQL endpoint RDF browser and the Chem2Bio2RDF faceted RDF browser http://chem2bio2rdf.org/exhibit/drugbank.html). Currently it provides the following views:

- Table view (see Figure 4a): The drug structure and related properties are displayed in a table format and the faceted filters on the right side can narrow down the view. For example, shown here are drugs that been approved by FDA in 2001 with the anti-allergic agent category;
- Timeline view (See Figure 4b): It displays the timeline of drugs based on the year they get approved by FDA;
- Tile view (see Figure 4c): It groups drugs first by year of approval, then by alphabetic order of drug names;
- Thumbnail view (See Figure 4d): It displays drugs based on their drug names with drug structures as thumbnails.

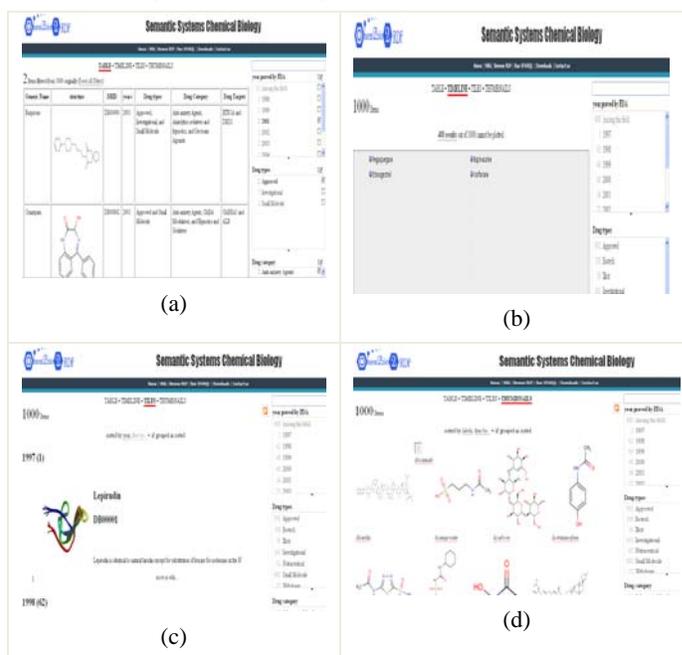

(a) (b)
(c) (d)

Figure 4. Screenshots of Chem2Bio2RDF Portal

## 3.3 Automatic SPARQL generator

If the users want to explore the relation between two classes, they do not necessarily need to know the specific dataset and the linkage between them, Link Path Generator (LPG) can automatically explores all possible link paths based on Systems Chemical Biology Data Source Ontology that models the semantic description of the datasets and their linkages. More specifically, LPG contains a graph mining module to enumerate all the possible routes from between two data sources within the

linked graph, which can be translated into concrete SPARQL queries easily. LPG is written as:

$$R(A,B) \xrightarrow{Ontology} R(\{A_1, A_1, \cdots A_m\}, \{B_1, B_1, \cdots B_n\}) \xrightarrow{network} \sum_i^m \sum_j^n path_{A_i \rightarrow B_j}$$

Where *A* and *B* are two classes which can be *Chemical*, *Biological*, *Pathway* and so on. *m* and *n* is the number of data sources relating to *A* and *B* respectively. The relation between *A* and *B* can be derived from the union of the paths between all the sources of *A* and all the sources of *B*. We implement a graph mining algorithm to enumerate all the unique, non-redundant (one that doesn't revisit a node multiple times) paths between $A_i$ and $B_j$. For example, if we are looking for relation between *Pathway* and *Side effect*, 14 paths are generated by LPG as listed below:

1) [sider -> kegg] contains 7 paths:

[sider, compound_hub, bindingdb_ligand, bindingdb_protein, uniprot_hub, kegg]
[sider, compound_hub, ctd, gene, gene2uniprot, uniprot_hub, kegg]
[sider, compound_hub, drugbank_drug, drugbank_target, uniprot_hub, kegg]
[sider, compound_hub, matador, uniprot_hub, kegg]
[sider, compound_hub, pubchem_bioassay, gi, gi2uniprot, uniprot_hub, kegg]
[sider, compound_hub, qsar, gene, gene2uniprot, uniprot_hub, kegg]
[sider, compound_hub, ttd_drug, ttd_target, uniprot_hub, kegg]

2) [sider -> reactome] contains 7 paths:

[sider, compound_hub, bindingdb_ligand, bindingdb_protein, uniprot_hub, reactome]
[sider, compound_hub, ctd, gene, gene2uniprot, uniprot_hub, reactome]
[sider, compound_hub, drugbank_drug, drugbank_target, uniprot_hub, reactome]
[sider, compound_hub, matador, uniprot_hub, reactome]
[sider, compound_hub, pubchem_bioassay, gi, gi2uniprot, uniprot_hub, reactome]
[sider, compound_hub, qsar, gene, gene2uniprot, uniprot_hub, reactome]
[sider, compound_hub, ttd_drug, ttd_target, uniprot_hub, reactome]

Based on the link path, SPARQL can be constructed, and then all the results are combined to output.

### 3.4 MEDLINE/PubMed Literature Cross-validation

MEDLINE/PubMed is the most frequently consulted online scientific medical resource in the world. In our Chem2Bio2RDF portal, we are processing 17,862,546 scientific abstracts from 1966 to present to rdf format. Not only the normal citation information, i.e. authors, journal, publication date, etc. have been processed, but also the Chem/Bio information have also been extracted. Current, our extractions are based on exact dictionary match. We used the PubChem as the dictionary to identify the compounds. The Uniprot human genes are used as the dictionary for the genes. The MESH disease terms and the COSTART side effects are applied to diseases and side effects extraction. As shown in Figure 5, the chemical compounds, genes, diseases and side effects for each abstracts are extracted and then associated with other data sources.

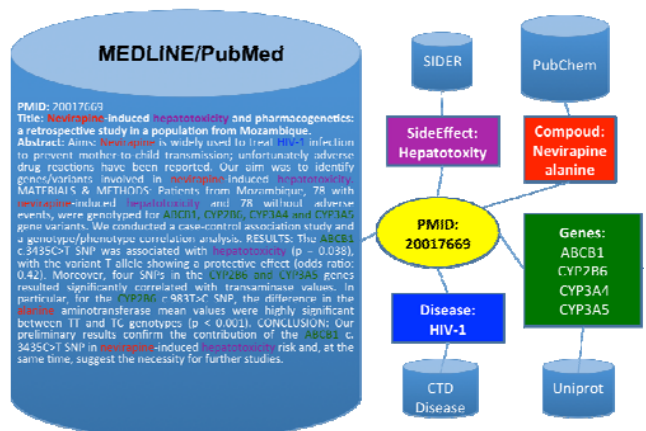

Figure 5, Chem/Bio Extraction for MEDLINE/Pubmed

After the Chem/Bio terms extraction, each literature is connected to other notes (chemical, genomic, phenotype, etc.) in the Chem2Bio2RDF graph, which provides a more meaningful way to search and classify the literature based on domain knowledge. For instance, if users want to study the literatures resource for a given drug, they can retrieve the literatures that contain either the given drug or the compound with a similarity value greater than a given threshold. If users are interested in studying certain disease, both the literatures contain the given disease and the literatures contain the compounds/genes that associated with the given disease can be retrieved. Those potential search methods enable a broad way to retrieve documents based on domain knowledge.

Meanwhile, the Chem/Bio information extraction for MEDLINE/PubMed provides a cross-validation for our Chem2Bio2RDF search result. As shown in Figure 6, the left window allows users to specify two search terms, for example, find the relationships for the compound, doxazosin, and the side effect, necrosis. The right window shows the network visualization based on our RDF data, which connect the compound with its targets, targets with pathways, and then associated the pathways with side effects. More details about this network validation are shown in case 2. The bottom window provides the MEDLINE/PubMed results for this given input by retrieving the literatures that contain both inputs and the associated genes/pathways. This MEDLINE/PubMed validation can also used to rank the generated paths based on literature confidence.

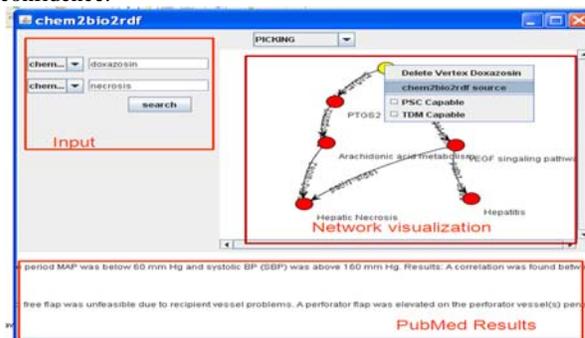

Figure 6. MEDLINE/PubMed cross-validation service

### 3.5 Cytoscape Plugin

Cytoscape is an open source bioinformatics tool originally developed for visualizing and analyzing biological networks. Over the past few years, Cytoscape has been used for the analysis of chemgenomics data which explores the relationships between compounds, genes and diseases. Since Cytoscape allows users to develop plugins that are best suited to the needs of their own dataset, we developed a plugin called Chem2Bio2RDFViz that allows users to interact with the Chem2Bio2RDF dataset in the most efficient manner possible (see Figure 8).

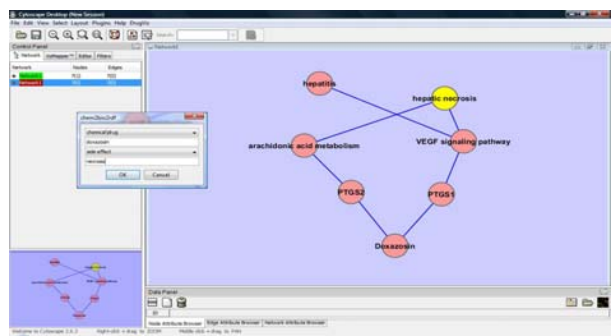

Figure 8. Cytoscape plugin prototype

Chem2Bio2RDFViz allows you to enter the names of the drug/protein/side effect/pathway/disease/gene whose network you would like to visualize and analyze. Once you enter the text, a network will be generated in the Cytoscape desktop which can be further manipulated according to the user's needs. Right clicking on each node leads to a range of choices that give you more information related to the network. The figure below shows the visualization of the network Doxazosin and its side effect Necrosis. Right clicking on the node Doxazosin will give you the MEDLINE/PubMed option that displays the MEDLINE/PubMed articles related to Doxazosin and Necrosis. Also, we can use the myriad inbuilt features of Cytoscape to understand the networks better.

## 4. Use cases

Here we provide three use cases to demonstrate the different levels of drug discovery.

## 4.1 Compound targets exploration

Given a drug, its targets are needed to be explored to understand its real mechanism, while for a drug-like compound, finding all its possible targets in the early discovery stage is desired in order to avoid the unexpected side effects in the clinical experiments. It's possible that many efforts have been made to identify the targets of a drug, but the results are scattered in the literature or hidden in some experimental records. This raises the following question:

**Question**: Given a drug (or chemical), find its possible targets (i.e., Gefitinib).

**SPARQL**:

*SELECT ?uniprot  WHERE {*
   *{?compound compound:CID ?compound_cid . FILTER (?compound_cid= 123631) .*
   *?chemical bindingdb_ligand:cid ?compound .*
   *?target bindingdb_interaction:monomerid ?chemical.*
   *?target bindingdb_interaction:uniprot?uniprot.*
   *?target bindingdb_interaction:ic50_value ?ic50 . FILTER (?ic50<10000) . }*
*UNION  {?compound compound:CID ?compound_cid . FILTER (?compound_cid= 123631) .*
   *?drug drugbank_drug:CID ?compound .*
   *?drugtarget drugbank_interaction:DBID ?drug.*
   *?drugtarget drugbank_interaction:human ?human .  FILTER (?human="1") .*
   *?drugtarget drugbank_interaction:SwissProt_ID ?uniprot.    }}*
*GROUP BY ?uniprot*

Gefitinib (CID=*123631*) is widely investigated in non-small cancer of the lung, colon cancer, breast cancer and cancer of the head and neck. We are only using DrugBank and BindingDB dataset to find its targets in this example. DrugBank offers approved drug targets from literature. As it considers non-human target, we used a filter to select only human targets. BindingDB provides drug target experimental interaction results. IC50 is a measurement of binding affinity, we selected ic50<10000nm (nm is the default unit) to guarantee the compound is able to interact the target.

While using only DrugBank, it yields one result P00533 (EGFR: Epidermal growth factor receptor), whose well known inhibitor is Gefitinib. While adding another dataset BindingDB, We got a new possible target P04626 (ERBB2: Receptor tyrosine-protein kinase erbB-2). EGFR (the known Gefitinib target) is part of the ERBB receptor family, which has four closely related members: EGFR (ERBB1), HER2 (ERBB2), HER3 (ERBB3) and HER4 (ERBB4), thus Gefitinib is active against ERBB2 is not surprising, however, DrugBank does not have this information, thus it's necessary to consider all the datasets as much as possible.

## 4.2 Disease specific chemical discovery

Drug discovery process generally starts with the disease target identification, followed by hits selection (active compounds against the disease target). Finding the potential chemicals is the main task in the early drug discovery stage. As so many public data emerges, it becomes possible to find some intriguing chemicals by running one SPARQL. This is particularly of interest to some lower-profitable disease like malaria that is paid little attention by pharmaceutical companies. But academia and the small companies do not have adequate funding to screen millions of compounds in order to find the hits. Chem2Bio2RDF attempts to collect all this kind of public data and the compounds can be easily obtained by the following SPARQL.

**Questions**: Find a disease (i.e., Malaria) specific chemicals
**SPARQL**:
*SELECT * WHERE {*
 *?chemogenomics chemogenomics:CID ?compound_cid .*
 *?chemogenomics chemogenomics:GENE ?gene_symbol .*
 *?omim omim:gene ?gene_symbol .*
 *?omim    omim:Disorder_name           ?disease    .   FILTER regex(?disease,"Malaria","i") .*
*}*

The SPARQL starts to find disease causing genes in OMIM and then select the associated compounds of the genes. 1,003 compounds and 5 disease causing genes are returned from this query, which enable scientists to do further experiments.

## 4.3 Adverse drug reaction

Adverse reaction in drug usage, as the notion self-suggests, has serious consequence and is often subject to rigorous investigation in pharmaceutical R&D processes. In this case, we integrate into Chem2Bio2RDF another scenario to study the most significant pathways that are associated with a given drug affect. The association between side effect and pathway is made possible using pathway's gene components that are targets of related drugs. More specifically, we consider a gene is related to a certain side effect if at least two drugs targeting this gene incur the side effect. On top of that, if there exists a pathway that contains more than 2 gene targets that associated with that side effect, an associative relationship between the pathway and side effect can be drawn. Clearly, the more these associative paths can be discovered, the stronger the evidence of such pathway-adverse drug effect association it becomes.

**Question**: Find the top 5 pathways in the KEGG pathway contain at least two of the efficient target that associated with a given side effect (i.e. *hepatomegaly*). A gene target is consider as efficient if the gene is targeted by at least two drugs that cause the given side effect.
**SPARQL**:
*SELECT ?pathway_id (count(?pathway_id) as ?count) WHERE {*
 *?sider2compound sider:side_effect ?side_effect . FILTER regex(?side_effect,"hepatomegaly","i") .*
 *?sider2compound sider:cid ?compound .*

 *?drug drugbank_drug:CID ?compound .*
 *?drug2target drugbank_interaction:DBID ?drug .*
 *?drug2target drugbank_interaction:SwissProt_ID ?uniprot .*
 *?kegg_pathway kegg_pathway_protein:Uniprot ?uniprot .*
 *?kegg_pathway kegg_pathway_protein:PathwayID ?pathway_id .*
 *} GROUP BY ?pathway_id ORDER BY ?count*

In this SPARQL, we need to first map side effect to drugs, and then map drug to targets. At the end, the target will be mapped to pathway. All the possible connections between side effect and pathway are counted to get the most reasonable pathways that associated with the given side effect.

**Results**: Available studies suggest that hepatic toxicity has been the most frequent single cause of safety-related drug withdrawal (e.g.,ticrynafen, benoxaprofen, bromfenac, troglitazone, nefazodone). Hepatotoxicity discovered after approval from marketing also has limited clinical use of many drugs, including isoniazid, labetalol, trovafloxacin, tolcapone, and felbarmate. Thus, it is important to systematically review the compounds, gene target and pathways that associated with liver injury.

In our mapping process, we use the hepatic necrosis, hepatitis and hepatomegaly as an example to study the relationship between drugs, targets, pathways and side effect. The graph is shown as following:

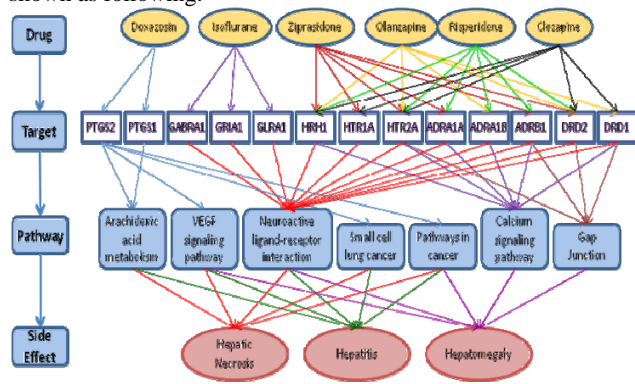

Figure 9: Adverse Drug Reaction

Figure 9 shows the top 5 pathways that associated with the three majority liver injury. It demonstrates that the mechanisms for hepatic necrosis and hepatitis are very close. They share the top 5 pathways: Arachidonic acid metabolism, VEGF signaling pathway, Neuroactive ligand-receptor interaction, small cell lung cancer, and pathways in cancer. The mechanism for hepatomegaly is a little bit different. The top 5 pathways of hepatomegaly contain the calcium signaling and gap junction pathway, which are not involved in the hepatic necrosis and hepatitis. Literature review [14] shows that those pathways are highly correlated with liver injury. For instance, the increase concentration of calcium in the calcium signaling pathway will cause the damage of hepatic cell. The targets we discovered are also known as the major targets for liver diseases based on literature reviews [11].

## 5. Evaluation

We evaluate the outcomes from the previous case studies against the confirmed evidence in literature search as well as salient domain knowledge. In addition, we also developed comprehensive assessment to examine the coverage of datasets in their associated domains. Finally we illustrate the improvement gained through Chem2Bio2RDF as opposed to our previous work where no semantic web framework is deployed.

### 5.1 Study of systems chemical biology

The difficulties of polypharmacology are to explore the combination of targets and then to identify active compounds against the sets of targets. Linking between chemical, biological, systems, and phenotype data is demonstrated to be a promising way to address the problems. For example, linking between bioassay data and market drug data enables to explore the compounds similar to the drug that already shows polypharmacology. Quinacrine, which has been used as an anthelmintic and in the treatment of giardiasis and malignant effusions, shows polypharmacology. One compound loxapine (CID 71399) is found to show similar polypharmacology with quinacrine. Loxapine is active in both BioAssay 859 and BioAssay 377, whose targets are CHRM1 and ABCB1 respectively. As loxapine tends to be hydrophobic molecules, medicinal chemists would not be surprised that it is active in BioAssay 377 which identifies substrates (or inhibitors) for multidrug resistance transporter. It is also reported that loxapine might get metabolized to amoxapine that is a considerably weak antagonist in BioAssay 859 [9]. Other than loxapine, many identified compounds such as oxybutynin and dexamethasone were proved to show polypharmacology by literature reviews.

While we link bioassay data to pathways, we could identify the compounds that inhibit at least two of proteins in a pathway, leading to the pathway dysfunction. For example, compound CID 6,419,769 could interact with proteins HSD11B1 and AKR1C4, which are in the different branches of C21-Steroid hormone metabolism pathways. The blocking of the pathway might be able to partially explain why CID 6,419,769 has side effect [1]. In protein-protein interaction network, two proteins are connected if both are physically interacted. In terms of polypharmacology, the deletion of one protein does not affect the whole network, but if two connected nodes with high degree were deleted, the network would be disturbed. For example, by linking bioassay to PPI, we found that two compound (CID 460,747 and CID 9,549,688) are active against two high degree proteins (PLK1 and TP53) which are associated with cancer. Via linking data sources among different domains, not only promising compounds to be high effective could be identified but also the risk of compounds could be somehow evaluated.

### 5.2 Dataset and result coverage

There are parallel contributions from different data sources and vendors toward same domain (for example, KEGG and Reactome are two independent data vendor provides same datum coverage over pathway domain), therefore the approach in which one single data source is made delegating the entire domain could produce incomplete outcome. To avoid this potential pitfall, for each domain in Chem2Bio2RDF, we have collected from a variety of data sources to make the ensemble as complete as possible. In particular, coverage for systems and chemogenomics is under close scrutiny due to their central roles in system chemical biology. Here we list the percentage coverage for PPI, pathway and chemogenomics and demonstrate the significance of integration using semantic web.

In PPI (Table 1), HPRD and DIP have 35,645 and 32,976 unique protein pairs respectively, and the total number of unique pairs in two datasets is 67,769. Each dataset contributes almost half of the pairs, and both share very little number of common pairs. The PPI network would not be complete if either dataset were ignored. Pathway is more complicated than PPI, since each organization could have its own definition of pathway, which makes the whole integration very difficult. For example, the pathway in Reactome is usually composed by a small number of proteins, although the total number of pathways is more than KEGG, the proteins involved in Reactome are far less than KEGG (Table 2). We are not able to judge which one is better, thus we have to consider all pathway datasets together. Figure 10 shows the dataset distribution of chemogenomics data. A chemical protein (gene) interaction is recorded as one entry, and all the unique interactions were derived from 10 datasets. We did not consider another two chemogenomics data sets (KEGG Ligand and PharmGKB), as KEGG Ligand includes only metabolic molecules rather than chemicals designed for drug discovery and many drugs in PharmGKB only provide names from which the chemical identifier is not able to be linked to compound. Many datasets only contribute a small portion of interactions so that it is not able to represent all chemogenomics data. Some of the datasets are quite small but they cannot be ignored. For example, BindingMOAD provides the protein and ligand complex crystal information which is the most accurate binding data. Kidb (called PDSP) is only interested in the receptors rather than all the protein spaces. DrugBank provides the most comprehensive drug and its target interactions.

Table 1: PPI data source distribution

| Data source | # of records | percentage |
|---|---|---|
| HPRD | 35645 | 52.6% |
| DIP | 32976 | 48.7% |
| ALL | 67769 | |

Table 2: Pathway data source distribution

| Data source | protein | | pathway | |
|---|---|---|---|---|
| | # of records | percentage | # of records | percentage |
| KEGG | 8172 | 81.0% | 192 | 34.8% |
| Reactome | 4397 | 43.6% | 360 | 65.2% |
| ALL | 10091 | | 552 | |

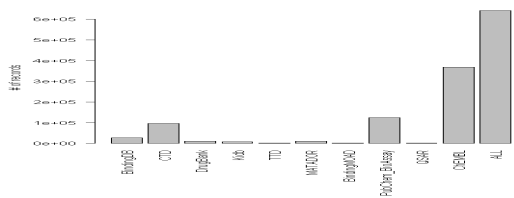

Figure 10: chemogenomics data source distribution

One dataset hardly be able to cover all the records of one domain, thus missing some of the datasets might neglect many link paths, sometimes, the results might even be changed significantly. We designed the following experiments to answer to the case 1 question, where we are searching for disease specific chemicals. In this task, we need to first identify disease genes (OMIM and PharmGKB) and then find chemicals interacting with the gene or gene expression products (BindingDB, DrugBank, ChEMBL and PubChem BioAssay). In our evaluation, we either selected one dataset in each step or selected all the datasets. As the table 4 shows, the number of chemicals retrieved from all datasets is far more than that only one dataset is considered.

Table 4: Results of discovering disease specific chemicals

| Dataset used | # of chemical discovered |
|---|---|
| OMIM, DrugBank | 77 |
| OMIM, BindingDB | 175 |
| OMIM, PubChem | 3577 |
| OMIM, ChEMBL | 2036 |
| PharmGKB,DrugBank | 292 |
| PharmGKB, BindingDB | 1256 |
| PharmGKB, PubChem | 28410 |
| PharmGKB, ChEMBL | 20513 |
| ALL | 45606 |

## 5.3 Comparison with previous work

In the previous study, only 4 datasets (PubChem, DrugBank, KEGG and HPRD) were studied [5], no attempt is made to integrate multiple datasets within one domain simultaneously. In Chem2Bio2RDF, efforts have been made to provide coverage over the comprehensive landscape of system chemical biology, 23 datasets so far has been integrated, as a direct result; the expected entity pairs discovered has increased significantly. Moreover, LPG would answer user's queries by automatically locating related datasets.

In addition, the performance has been improved as well as lots of time was required to parse the heterogeneous datasets previously. But now, since data sources have been published as RDF triples, SPARQL queries can be used to directly identify hidden knowledge.

In the work of adverse drug reaction study[22], it relies on the internal data source which includes 1,458,680 unique compounds and 2,190 unique targets, Chem2Bio2RDF currently covers 235,313 compounds, 19,534 genes, 641, 855 chemogenomics which are public accessible. However, we must take care of the quality of data in public domain. We also need to build robust chemogenomics predictive model based on the rich chemogenomics data source, as it is impossible that every compound is tested against all the targets.

## 6. Conclusion and Future work

This paper discusses our on-going effort to create an user-friendly Linked Open Data portal for chemical biology. It applies semantic web technologies to integrate data and identify hidden association. Comparing with other LOD portals, currently Chem2Bio2RDF portal contains the following features: a RDF faceted browser, an automatic SPARQL query generator, a MEDLINE/PubMed literature cross-validation service, a cytoscape pulgin, and a WENDI 2.0 service for disease-drug-compound discovery.

Our future work will focus on: (1) extending cytoscape visualization plugin to enable analytical graph mining and association detection; (2) adding a predictive model for compound gene associations, as not every compound is tested against all the targets, a predictive model is necessary; (3) applying CRF(common random field) name entitiy method to identify the chem./bio terms to provide an extentable gene, compound, pathway, disease and drug extraction for MEDLINE/PubMed literatures; (4) enriching provenance data to the current Chem2Bio2RDF data to enable provenance-based filtering and visualization; and (5) identifying semantic associations among different types of entities to enable complex drug discovery.

### Acknowledgement
This work is funded by NIH VIVO project (UF09179) and Eli Lilly.